\documentclass[prl,preprint,aps]{revtex4} \usepackage{graphicx}
\begin{document} \title{Quasi-continuous atom laser in the
presence of gravity} \author{F. Gerbier, P. Bouyer and A. Aspect}
\affiliation{Groupe d'Optique Atomique, Laboratoire Charles Fabry
de l'Institut d'Optique, UMRA 8501 du CNRS, B\^{a}t. 503, Campus
universitaire d'Orsay, B.P. 147, F-91403 ORSAY CEDEX, FRANCE}
\begin{abstract} We analyse the extraction and the subsequent fall
of a coherent atomic beam from a trapped Bose-Einstein condensate
using a rf transition to a non-trapping state, at T=0 K. Our
treatment fully takes gravity into account but neglects the
details of the mean-field potential exerted on the free falling
beam by the trapped atoms. We focus on the weak coupling regime,
where analytical expressions for the output rate and the output
mode of the ``atom laser'', can be obtained. Comparison with
experimental data of Bloch {\it et al. }[Phys. Rev. Lett. {\bf
82}, 3008 (1999)] without any adjustable parameter is
satisfactory.
\end{abstract} 
\date{\today} \maketitle

Bose-Einstein condensates (BEC) of dilute alkali vapors
\cite{firstBEC} constitute a potential source of matter waves for
atom interferometry,  since it has been proven \cite{andrews} that
they are inherently fist-order coherent. Various schemes for
``atom lasers'' have been used to extract a coherent matter wave
out of a trapped BEC. Pulsed devices \cite{mewes,hagley,anderson}
were demonstrated by using an intense spin-flip radio frequency
(rf) pulse, Raman transitions or gravity-induced tunneling from an
optically trapped BEC. Later on, a quasi-continuous atom laser has
been demonstrated by using a weak rf field that continuously
couples atoms into a free falling state \cite{bloch}. This
``quasi-continuous" atom laser promises spectacular improvements
in application of atom optics, for example in the performances of
atom-interferometer-based inertial sensors \cite{bouyer}.\\
Gravity plays a crucial role in outcouplers with spin-flip rf
transitions: it determines the direction of propagation of the
extracted matter wave, its amplitude and its phase. However, most
of the theoretical studies do not take it into account (for an
up-to date review, see Ref.\cite{ballagh2}). Although gravity has
been included in numerical treatments \cite{zhang} relevant for
the pulsed atom laser of Mewes {\it et al.} \cite{mewes}, to our
knowledge, only the 1D simulations of
Refs.\cite{ballagh2,schneider} treat the quasi-continuous case in
presence of gravity, and the results of \cite{schneider} compare
only qualitatively to the experimental data of \cite{bloch}. \\ In
this Letter, we present a 3D analytical treatment fully taking
gravity into account. The extraction of the atoms from the trapped
BEC and their subsequent propagation under gravity are
analytically treated in the weak coupling regime relevant to the
quasi-continuous atom laser. We derive an expression for the atom
laser wave function and a generalized rate equation for the
trapped atoms that agrees quantitatively with the experimental
results of \cite{bloch}, without any adjustable parameter.\\
We consider a $\mathrm{^{87}Rb}$ BEC in the $F=1$ hyperfine level
at $T=0$ K. The $m=$-1 state is confined in a harmonic magnetic
potential $ V_{\mathrm{trap}}=\frac{1}{2}M(\omega_{x}^{2} x^{2}
+\omega_{\perp}^{2} y^{2} +\omega_{\perp}^{2} z^{2})$. A rf
magnetic field ${\bf B_{\rm rf}}= B_{\rm rf} \cos{( \omega_{\rm
rf} t)} {\mathbf{e_{x}}}$ can induce transitions to $m=$0
(non-trapping state) and $m=$+1 (expelling state). The condensate
three-component spinor wavefunction ${\bf \Psi^{'}}= \lbrack
\psi^{'}_{m} \rbrack_{m=-1,0,+1}$ obeys a set of coupled
non-linear Schr\"{o}dinger equations \cite{ballagh1}. Following
Steck {\it{et al.}} \cite{steck}, we consider the ``weak coupling
limit''. In this regime to be defined more precisely later, the
coupling strength is low enough that the populations $N_{m}$ of
the three Zeeman sublevels obey the following inequality: $N_{+1}
\ll N_{0} \ll N_{-1}$. In the rest of this paper, we therefore
restrict ourselves to $m=$-1 and $m=$0, and set the total atomic
density $n({\mathbf{r}}) \approx \mid \psi_{-1}( {\mathbf{r}},t)
\mid^{2}$. \\ At this stage, the components $\psi_{m} =
\psi_{m}{'} e^{-i m \omega_{ \mathrm{rf}} t}$ obey, under the
rotating wave approximation, the following two coupled equations:
\begin{mathletters} \begin{eqnarray} \label{gp1}  i \hbar
\frac{\partial \psi_{-1}}{\partial t} & = & [\hbar
\delta_{\mathrm{rf}} + \mathcal{H}_{-1} ] \psi_{-1}+ \frac {\hbar
\Omega_{\mathrm{rf}}}{2} \psi_{0}\\ \label{gp2} i \hbar
\frac{\partial \psi_{0}}{\partial t} \ & = & \mathcal{H}_{0}
\psi_{0} \ +  \ \frac {\hbar \Omega_{\mathrm{rf}}}{2} \psi_{-1}
\end{eqnarray} \end{mathletters}
with $\mathcal{H}_{-1} ={\mathbf{p}}^{2} / 2 M + V_{\mathrm{trap}}
+ U \mid \psi_{-1} \mid^{2}$ and $\mathcal{H}_{0}
={\mathbf{p}}^{2}/2 M-Mgz + U \mid \psi_{-1} \mid^{2}$. The
strength of interactions is fixed by $U=4 \pi \hbar^{2} a N/ M$,
$a \approx 5$ nm being the diffusion length that we suppose equal
for all collision processes, and $N$ the initial number of trapped
atoms. The rf outcoupler is described by the detuning from the
bottom of the trap  $\hbar \delta_{\mathrm{rf}}= V_{{\rm {off}}} -
\hbar \omega_{\rm{rf}}$ and the Rabi frequency $\hbar
\Omega_{\mathrm{rf}}= \mu_{\mathrm{B}}B_{ \mathrm{rf}} / 2
\sqrt{2}$ (taking the Land\'e factor $g_{\mathrm{F}}$=-1/2). The
origin of the $z$ axis is at the center of the condensate,
displaced by gravity from the trap center by $z_{{\rm
sag}}=g/\omega_{\perp}^{2}$. We have taken the zero of energy at
$z=0$ in $m=$0, so that the level splitting at the bottom of the
trap is $V_{{\rm off}}= \mu_{\mathrm{B}}B_{0}/2 + Mg^{2} / 2
\omega_{\perp}^{2}$ ($B_{0}$ is the bias field). \\ The evolution
of the $m=$-1 sublevel at $T=0$ K is described \cite{dalfovo} by a
decomposition of the wave function into a condensed part and an
orthogonal one that describes elementary excitations
(quasiparticles). As $\mathcal{H}_{-1}$ depends on time through
$\mid \psi_{-1} \mid$, both the coefficients and the eigenmodes
depend on time. However, we assume an adiabatic evolution of the
trapped BEC \cite{choi}, so that the uncondensed component remains
negligible and:
\begin{equation} \label{trappedbec} \psi_{\mathrm{-1}}({\bf{r}},t)
\approx \alpha(t) \phi_{\mathrm{-1}}({\bf {r}},t) e^{-i
\int_{0}^{t}(\mu(t^{'})/ \hbar + \delta_{\rm rf}) dt^{'}}
\end{equation} where $\alpha(t)$=$(N_{-1}(t)/N)^{1/2}$. The time-dependent ground state
$\phi_{\mathrm{-1}}$, of energy $ \mu$ is, in the Thomas-Fermi
(TF) approximation \cite{dalfovo}, $\phi_{\mathrm{-1}} ({\bf
r},t)$=$(\mu / U)^{1/2}
[1-\tilde{r}_{\perp}^2-\tilde{z}^2]^{1/2}$, where
$\tilde{r}_{\perp}^2$=$(x/x_{0})^2+(y/y_{0})^2$ and
$\tilde{z}$=$z/z_{0}$ are such that $\tilde{r}_{\perp}^{2} +
\tilde{z}^{2} \leq 1$. The BEC dimensions are respectively
$x_{0}$=$(2 \mu / M \omega_{x}^{2})^{1/2}$ and $y_{0}$=$z_{0}$=$(2
\mu / M \omega_{\perp}^2)^{1/2}$. The condensate energy is given
by $\mu$=$(\hbar \omega/2)(15 a N_{-1} / \sigma)^{2/5}$ (we have
set $\omega$=$(\omega_{x} \omega_{\perp}^{2})^{1/3}$ and
$\sigma$=$(\hbar/M \omega)^{1/2}$). The time-dependence of
$\phi_{-1}$ is contained in $\mu$, $x_{0}$, $y_{0}$ and $z_{0}$,
which decrease with $N_{-1}$(t). In the following, we will take
typical values corresponding to the situation of \cite{bloch}
(reported in \cite{schneider}): $N=7 \times 10^{5}$ atoms
initially, $\omega_{x}$=$2\pi\times 13$ Hz and
$\omega_{\perp}$=$2\pi\times 140$ Hz, which gives $x_{0} \approx
55.7 \mu$m, $z_{0} \approx 5.3 \mu$m and $\mu/h \approx 2.2$ kHz.
\\ For the $m$=$0$ state, the hamiltonian of equation (\ref{gp2})
with $\Omega_{{\rm rf}}=$0 becomes in the TF approximation
\cite{edwards}: $\mathcal{H}_{0}$ =${\mathbf{p}}^{2}/2 M-Mgz + \mu
{\rm Max}[0,1-\tilde{r}_{\perp}^2-\tilde{z}^2]$. There is a clear
hierarchy in the energy scales associated with each term in
$\mathcal{H}_{0}$. First, in the TF regime, the kinetic energy in
the trap is negligible as compared to the mean field energy.
Second, for the atom numbers involved in current experimental
setups, $\mu / Mgz_{0}$ is small, and the mean field term is only
a perturbation as compared with gravity. The leading term is
therefore the gravitational potential, that subsequently converts
into kinetic energy along $z$. On the contrary, the
{\it{transverse}} kinetic energy (along $x$ and $y$) remains small
(we will make use of this to compute the output rate). In
addition, we neglect for simplicity the spatial variations of the
mean field potential: we replace it by the mean-field contribution
to the energy $\mu$ needed to remove one atom, {\it{i.e.}}
$E_{{\mathrm{mf}}}=4 \mu /7$ \cite{hagley}. In summary, we
consider in the following the approximated hamiltonian
$H_{0}\approx{\mathbf{p}}^{2}/2 M-Mgz + 4\mu/7$.\\ The eigenstates
of $H_{0}$ are factorized products of one dimensional
wavefunctions along the three axis. In the horizontal plane (x,y),
the eigenstates are plane waves with wavevectors $k_{x}$,$k_{y}$
that we quantize with periodic boundaries conditions in a 2D box
of size L. Consequently, the wavefunction is
$\phi_{0}^{\perp}(x,y)= L^{-1} e^{i(k_{x}x + k_{y}y)}$ and the
density of states $\rho_{xy}=L^{2} / 4 \pi^{2}$. Along the
vertical direction z, the normalizable solution of the 1D
Schr\"{o}dinger equation in a gravitational field is
\cite{landaumq} $\phi_{0}^{(z_{E_{z}})}$= $A.Ai(-\zeta_{E_{z}})$,
where $Ai$ is the Airy function of the first kind and
$\zeta_{E_{z}}=(z-z_{E_{z}})/l$. The classical turning point
$z_{E_{z}}=-E_{z}/Mg$ associated with the vertical energy $E_{z}$
labels the vertical solution; $l=(\hbar^{2} / 2M^{2}g)^{1/3}$ is a
length scale, such that $l \ll x_{0},y_{0},z_{0}$ (for $^{87}Rb$,
$l \approx 0.28 \mu$m). In order to normalize $Ai$ and to work out
the density of states, we restrict the wavefunction to the domain
$\rbrack -\infty,H \rbrack$, where the boundary at $z$=$H$ can be
arbitrarily far from the origin. In $\rbrack -\infty,z_{E_{z}}
\rbrack$,  $Ai$ falls off exponentially over a distance $l$, while
it can be identified with its asymptotic form $Ai(-s) \approx
\pi^{1/2} s^{-1/4} \cos(2s^{3/2}/3-\pi/4)$ for $s \gtrsim 0$. To
leading order in $H$, by averaging the fast oscillating $\cos^{2}$
term to $1/2$, we obtain the normalization factor $A$=$(\pi
H^{-1/2}/l)^{1/2}$. The longitudinal density of states follows
from $\rho_{z}(E_{z})=h^{-1} d \mathcal{V}/ dE$, where
$\mathcal{V} = \int_{\epsilon \leq E_{z}} dz dp_{z}
\theta(p_{z}^{2}/2M - Mgz-E_{z})$ (with $\theta$ the step
function) is the volume in phase space associated with energies
lower than $E_{z}$ . We restrict the $z$ space to the domain
defined above, and the $p_{\rm{z}}$ space to $p_{\rm{z}} \geq 0$.
In this way, we do not count the component of $Ai$ that propagates
opposite to gravity\cite{component}. We obtain $\rho
_{z}(z_{E_{z}})$=$(1 /2 \pi l) H^{1/2} $. Finally, the output
modes are given by $\phi_{0}^{(n)}({\bf r
})$=$\phi_{0}^{\perp}(x,y) \phi_{0}^{(z_{E_{z}})}(z)$, where $n$
stands for the quantum numbers $(k_{x},k_{y},z_{E_{z}})$, and the
density of modes is $\rho_{3D}^{(n)}$=$(1/8 \pi^{3}) L^{2} H^{1/2}
/l $.\\ Thus, the problem is reduced to the coupling of an
initially populated bound state $m$=$-1$, of energy
$E_{-1}$=$\hbar \delta_{rf} + \mu$ to a quasi-continuum of final
states $m$=$0$, with a total energy
$E_{0}^{(n)}$=$\hbar^{2}(k_{x}^2 + k_{y}^2)/2M - M g z_{E_{z}}
+4\mu/7$. A crucial feature in this problem is the resonant
bell-shape of the coupling matrix element $W_{\rm fi}$=$(\hbar
\Omega_{\rm rf}/2) \langle \phi_{0}^{(n)} \mid \phi_{\mathrm{-1}}
\rangle$. We work out the overlap integral $I^{(n)}$=$\langle
\phi_{0}^{(n)} \mid \phi_{\mathrm{-1}} \rangle$ in reduced
cylindrical coordinates $(\tilde{r_{\perp}},\theta,\tilde{z})$ and
set $\tilde{k}^{2}=(k_{x} x_{0})^2 +  (k_{y} y_{0})^2$. We
integrate over $\theta$ and $\tilde{r_{\perp}}$, and transform the
sum over z with the Parseval relation \cite{castin} to obtain: $
I^{(n)} = 2 \pi A x_{0} y_{0}/L (\mu/U)^{(1/2)}
\int_{-\infty}^{\infty} \tilde{g}(\tilde{k},v
)e^{i(z_{0}/l)(v^{3}/3 -v z_{E_{z}}/l)} dv$. In this integral,
$\tilde{g}$ is the Fourier transform with respect to $z$ of
$g(\tilde{k}, \tilde{z})=(p \cos{p}-\sin{p})/\tilde{k}^3$, with
$p(\tilde{k},\tilde{z})=\tilde{k} (1-\tilde{z}^2)^{1/2}$, and $(2
\pi)^{(-1/2)} e^{i(v^{3}/3 -v z_{E_{z}}/l)}$ the Fourier transform
of $Ai$. Since $z_{0}/l \gg 1$, the integrand averages to zero out
of a small neighbourhood of the origin, where the linear term in
$v$ is dominant. We obtain in this way: {\small{\begin{equation}
I^{(n)} = 2 \pi \frac{A l}{L} \sqrt{\frac{\mu}{U}}x_{0} y_{0} g(
\tilde{k} , \tilde{z_{E_{z}}})
\end{equation}}}
This overlap integral is non-vanishing only if the accessible
final energies $E_{0}^{(n)} \approx E_{-1}$ are restricted to an
interval $\lbrack -\Delta/2, \Delta/2 \rbrack$, where $\Delta$
$\sim$ $2Mgz_{0} \sim 22.7$ kHz. This gives a resonance condition
for the frequency $\omega_{\rm rf}$ \cite{bloch,choi}:
{\begin{equation} \mid \hbar \delta_{\rm rf} \mid \lesssim Mgz_{0}
\end{equation}
Two different behaviour can be expected in such a situation
\cite{choi,cohen}. In the strong coupling regime ($h \Gamma \gg
\Delta$, where $\Gamma$ is the decay rate worked out with the
Fermi Golden Rule), Rabi oscillations occur between the BEC level
and the narrow-band continuum. This describes the pulsed atom
laser experimentally realized by Mewes {\it{et al.}} \cite{mewes}.
On the contrary, in the weak coupling limit ($h \Gamma \ll
\Delta$), oscillations persist only for $t \leq t_{c}$=$h /
\Delta$, while for $t \geq t_{c}$, the evolution of the BEC level
is a monotonous decay. We have numerically verified this behaviour
on a 1D simulation analogous to \cite{schneider} (see
Fig.(\ref{Fig01}a)).

Eq.(\ref{gp2}) can be formally integrated
\cite{choi} with the help of the propagator $G_{0}$ of $H_{0}$: $
\Psi_{0}({\bf{r}},t) = \frac{\Omega_{\mathrm{rf}}}{2i}
\int_{0}^{t} dt^{'} \int d^{3}{\mathbf{r^{'}}}
G_{0}({\mathbf{r}},t;{\mathbf{r^{'}}},t^{'}) \Psi_{-1}
({\mathbf{r^{'}}},t^{'})$. Together with Eq.(\ref{gp1}) and
Eq.(\ref{trappedbec}), we can derive an exact integro-differential
equation on $\alpha^{2}$, the fraction of atoms remaining in the
BEC. In the weak coupling regime, the condition $h \Gamma
\ll\Delta$ expresses that the memory time of the continuum $t_{C}$
is much shorter than the damping time of the condensate level.
This allows one to make a Born-markov approximation
\cite{choi,jack}, which yields the rate equation: {\small
\begin{equation} \label{rate} \frac{d N_{-1}}{dt} = - \Gamma
(N_{-1}) \ N_{-1} \end{equation}} The output rate $\Gamma$ is
given by the Fermi golden rule:{\small{\begin{equation}
\label{Gamma} \frac{\Gamma}{\Omega_{\mathrm{rf}}^{2}} \approx
\frac{15 \pi}{32} \frac{ \hbar}{\Delta}  \mathrm{Max} [0,1 -
\eta^{2}]^{2}
\end{equation}}}
with $\eta$=$2(\hbar \delta_{\rm{rf}} + 4 \mu/7)/\Delta$. The rate
equation (\ref{rate}) is non linear, since $\Gamma$ depends on
$N_{-1}$ through $\Delta=2Mgz_{0}$. To obtain Eq.(\ref{Gamma}), we
have neglected the transverse kinetic energy as compared to
$E_{-1}$, and used the relations $\mu / U x_{0} y_{0} z_{0} = 15/8
\pi$ (valid in the TF approximation) and $\int_{0}^{+\infty} dw
 \mid w \cos{(w)} - \sin{(w)} \mid^{2}w^{-5}$=$1/4$. A
quasi-continuous output corresponds to the weak coupling regime
($h \Gamma \ll \Delta$). From Eq.(\ref{Gamma}), we deduce a
critical Rabi frequency $\Omega_{\rm rf}^{C} \sim 0.8 \Delta /
\hbar$ for which $h \Gamma \sim \Delta$. However, we have assumed
from the beginning that the BEC decay was adiabatic. This requires
$\mid
\partial h_{-1}/\partial t \mid \sim \Gamma \mu \ll
\epsilon_{(i)}^{2} / \hbar$, where $\epsilon_{(i)}$ is the energy
of the {\it{i}}th quasiparticle level in the trap. Taking
$\epsilon_{(i)} \gtrsim \hbar \omega_{\perp} $, we deduce from
Eq.(\ref{Gamma}) a condition on the Rabi frequency:
$\Omega_{\mathrm{rf}} \ll 1.6 ( g / z_{0})^{1/2}$. This condition
turns out to be much more stringent than the condition for weak
coupling $\Omega_{\mathrm{rf}} \ll \Omega^{C}_{\mathrm{rf}}$.
\\To compare our model to the data of Bloch {\it{et al}} \cite{bloch}, we have numerically integrated
Eq.(\ref{rate}) with the output rate (\ref{Gamma}) and their
experimental parameters. We show in Fig.(\ref{Fig02}) the number
of atoms remaining in the condensate after a fixed time as a
function of the detuning. The agreement with the experimental data
for the $\mid F=1;m_{F}=-1 \rangle$ state is good
(Fig.(\ref{Fig02}a)). To treat the $\mid F=2;m_{F}=2 \rangle$
case, the second trapping state $\mid F=2;m_{F}=1\rangle$ must be
included. We use the result of \cite{cohen} for the situation
where two discrete levels are mutually coupled, and only one of
them is also coupled to a continuum (with a decay rate
$\Gamma_{2,1}) $\cite{cohen}. For near resonant coupling, the
upper level acquires a decay rate $\Gamma_{2,2} \sim
\Gamma_{2,1}/2$ if $\Omega_{\rm rf} \gg \Gamma_{2,1}$. We have
taken this value as a first approximation. This leads to an
already good agreement with the experimental data
(Fig.\ref{Fig02}b), although better refinements is needed to
describe the full dynamics.\\ Finally, we want to point out that,
since $h \Gamma \ll \Delta$, the spatial region where outcoupling
takes place, of vertical extension $\delta z \sim h \Gamma / mg$,
is very thin compared to the BEC size. It allows us to think of
the outcoupling process in a semi classical way, in analogy with a
Franck-Condon principle \cite{band}: the coupling happens at the
turning point of the classical trajectory of the free falling
atoms. Neglecting the transverse kinetic energy in the calculation
of $\Gamma$ amounts to approximate the Franck-Condon surfaces by
planes. Since over the size of the BEC, the surface curvature is
small, the deviation from this plane is negligible compared to
$z_{0}$.\\ We have analyzed so far the extraction of the atoms
from the trapped BEC. We address now the question of the
propagation of the outcoupled atoms under gravity. For times $t
\gg t_{c}$, the explicit expression of the propagator $G_{0}$ of
$H_{0}$ is \cite{feynman}: {\small
\begin{equation} \label{propagator} G_{0}({\mathbf{r}},
{\mathbf{r}^{'}},\tau=t-t^{'})=(\frac{M}{2 \pi \hbar \tau})^{3/2}
e^{i S_{\mathrm{class}}({\mathbf{r}},
{\mathbf{r}^{'}},\tau=t-t^{'})} \theta(\tau)
\end{equation}}
Here, the classical action is given by
$S_{\mathrm{class}}({\mathbf{r}},
{\mathbf{r}^{'}},\tau=t-t^{'})=(M/2 \hbar \tau)(({\mathbf{r}} -
{\mathbf{r^{'}}})^2 + g \tau^2(z+z^{'})-g^2 \tau^4/12) + 4 \mu
\tau / 7\hbar$ and $\theta$ is the step function. We compute the
output wavefunction $\Psi_{0}$ with stationary phase
approximations. This amounts to propagate the wavefunction along
the classical trajectory: in the limit where $S_{\mathrm{class}}
\gg \hbar$(this is true as soon as the atom has fallen from a
height $l$), this is the Feynman path that essentially
contributes. We obtain the following expression for the outcoupled
atomic wave function, {\it{i.e.}} the atom laser mode
\cite{Borde}(see Fig.(\ref{Fig01}b)): {\small{\begin{equation}
\psi_{0}({\bf r},t) \approx -\sqrt{\pi } \frac{\hbar \Omega_{\rm
rf}}{ M g l} \phi_{\mathrm{-1}}(x,y,z_{\rm{res}})
\frac{e^{i\frac{2}{3} \mid \zeta_{\rm{res}} \mid^{3/2}-i
\frac{E_{-1}t}{ \hbar}}}{\sqrt{ \mid \zeta_{\rm{res}} \mid^{1/2}}}
F(t,t_{\rm{fall}})\label{wavefunction}
\end{equation}}} In this expression, $z_{\rm{res}}$=$\eta z_{0}/2$ is the point of
extraction, $t_{\rm{fall}}$=$(2/g)^{1/2}(z-z_{\rm{res}})^{1/2}$ is
the time of fall from this point,
$\zeta_{\rm{res}}$=$(z-z_{\rm{res}})/l$, and
$F(\beta)$=$\alpha(\beta) M(\beta)$: $M$ is the top hat function,
that describes the finite extent of the atom laser due to the
finite coupling time. We can deduce from Eq.(\ref{wavefunction})
the size of the laser beam in the $x$ direction $x_{{\rm out}}$ =
$x_{0}(1-(2 E_{-1} / \Delta)^2)^{1/2}$, and a similar formula for
$y_{\rm out}$. Because the semi-classical approximation neglects
the quantum velocity spread due to the spatial confinement of the
trapped BEC, this wavefunction has plane wave fronts. This
property will not persist beyond a distance $z_{R} \approx y_{\rm
out}^{4}/4l^{3}\sim$ a few mm, analog to the Rayleigh length in
photonic laser beams, where the transverse diffraction $(\hbar/M
y_{{\rm out}})(2z_{R}/g)^{1/2}$ becomes comparable to the size
$y_{\rm out}$. Beyond $z_{R}$, diffraction has to be taken into
account. However, this discussion neglects the (weak) transverse
acceleration provided by the repulsive mean field potential that
is likely to affect the transverse profile.
\\ In conclusion, we have obtained analytical expressions for the
output rate and the output mode of a quasi-continuous atom laser
based on rf outcoupling from a trapped BEC. Our treatment, which
fully takes the gravity and the 3D geometry into account, leads to
a good agreement with the experimental results of
Ref.\cite{bloch}. This points out the crucial role played by
gravity in the atom laser behaviour. More elaborated treatments
should take the interaction energy between the trapped BEC and the
atom laser, as well as finite temperature effects \cite{choi} into
account. Nevertheless, we believe that the present
zero-temperature theory is a valuable starting point to describe
experiments with the atom laser.
\\ {\bf Acknowledgements:} We would like to thank the members of
the IOTA Atom optics group for stimulating discussions, in
particular the BEC team: G. Delannoy, Y. Le Coq, S. Rangwala, S.
Richard, J. Thywissen and S. Seidelin. We are also grateful to Y.
Castin for the calculation of the overlap integral and to J.
Dalibard for his valuable comments. This work is supported by DGA
(contract 99 34 050) and EU (contract HPRN-CT-2000-00125). FG
acknowledges support from CNRS.

\begin{figure}[h]
\caption{Atom laser in the weak coupling regime ($\Omega_{\rm 
rf}=300$ Hz). 1a : time evolution 
of the laser intensity starting with $\sim 2000$ atoms. For $t\gg t_{c}$, the numerical integration  
of Eq.(1) agrees with the output rate of Eq.(\ref{Gamma}). 1b : 
Spatial intensity profile of the atom laser at $t\gg t_{c}$ according to 
Eq.(\ref{wavefunction})} \label{Fig01}
\end{figure}

\begin{figure}[t]
\caption{Number of trapped atoms after 20 ms of rf outcoupling,
starting with $\approx 7.2\times 10^{5}$ condensed atoms, $\Omega_{{\rm
rf}}$=312 Hz for the $\mid1;-1\rangle $ sublevel (left), and
$\approx 7.0\times 10^{5}$ atoms, $\Omega_{{\rm rf}}$=700 Hz for the
$\mid 2;2\rangle$ sublevel (right). Diamonds are the experimental
points from Bloch {\it{et al.}}, solid line is the prediction
based upon our model using the experimental parameters. 
Theoretical and experimental curves have
been shifted in frequency to match each other, since $V_{{\rm
off}}$ is not experimentally known precisely enough (a precision
of $10^{-3}$ Gauss for the bias field $B_{0}$ is required to know
$V_{{\rm off}}$ within a kHz uncertainty). } \label{Fig02}
\end{figure}

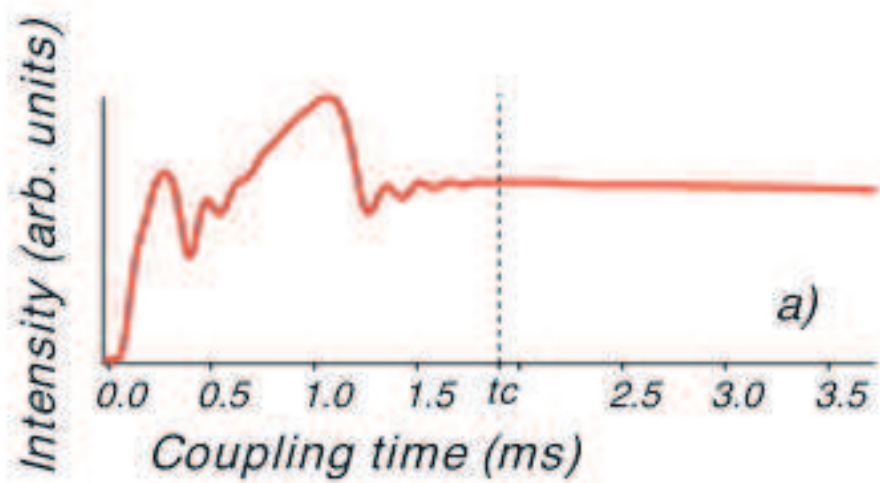
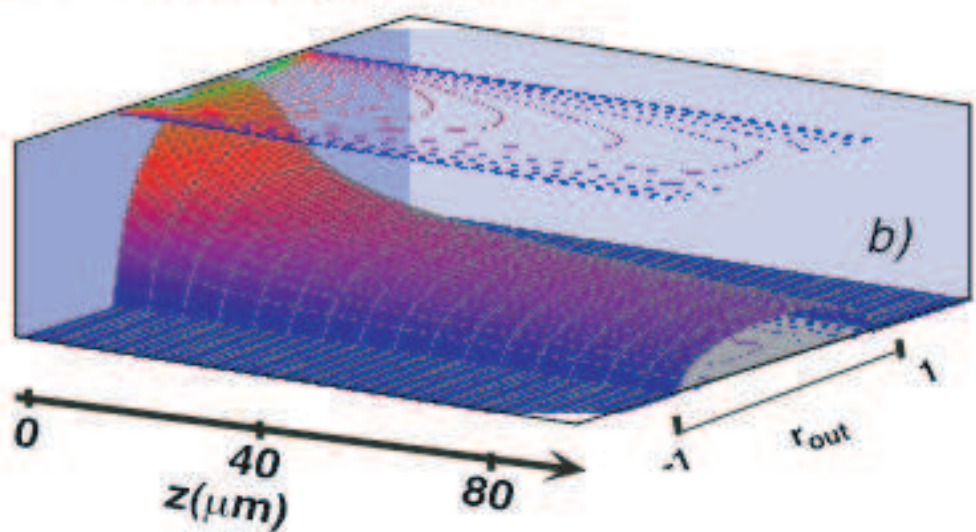
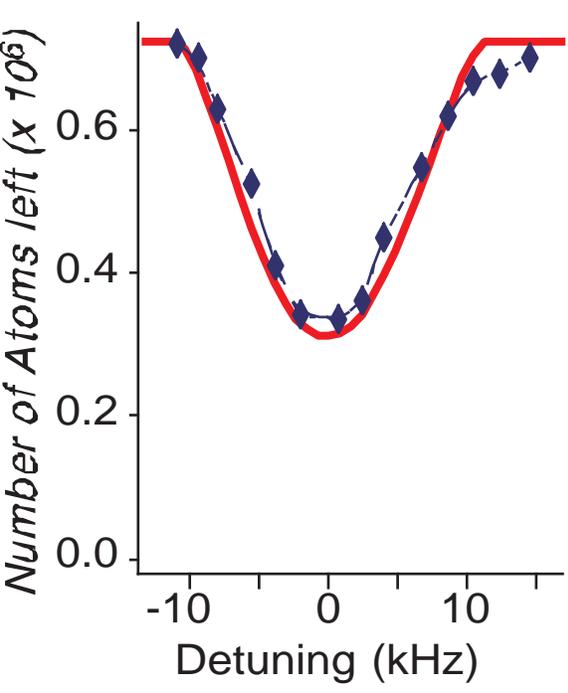
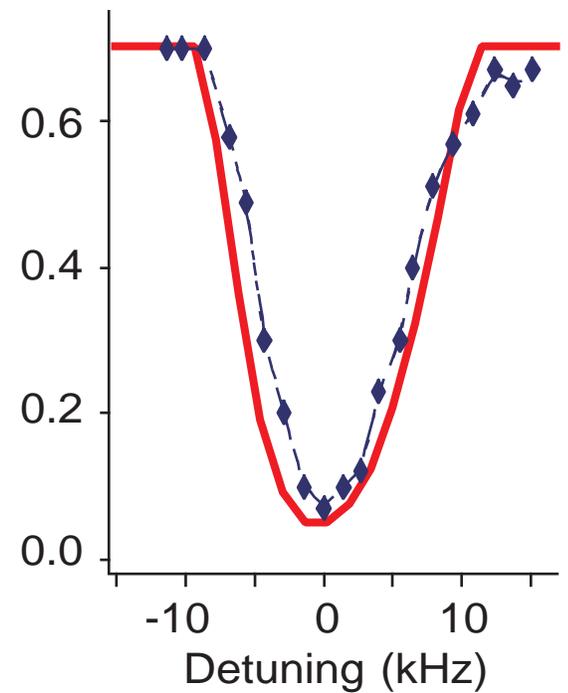
\end{document}